\documentclass[twoside,a4paper,11pt]{sea9}
\usepackage{graphicx}
\usepackage{hyperref}
\usepackage{movie15}
\begin{document}
\pagenumbering{arabic}
\pagestyle{myheadings}
\thispagestyle{empty}
{\flushleft\includegraphics[width=\textwidth,viewport=58 650 590 680]{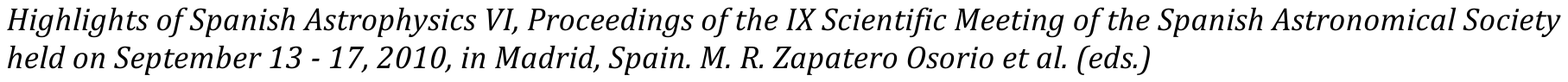}}
\vspace*{0.2cm}
\begin{flushleft}
{\bf {\LARGE
Morphogical Characteristics of OB Spectra and Environments}\\
\vspace*{1cm}
Nolan R. Walborn$^{1}$}\\
\vspace*{0.5cm}
$^{1}$
Space Telescope Science Institute, Baltimore, Maryland, USA\\
\end{flushleft}
%
\markboth{OB Spectra \& Environments}
{N.R. Walborn}
\thispagestyle{empty}
\vspace*{0.4cm}
\begin{minipage}[l]{0.09\textwidth}
\ 
\end{minipage}
\begin{minipage}[r]{0.9\textwidth}
\vspace{1cm}
\section*{Abstract}{\small
Following a brief exposition of the morphological process, two examples
of its application in contemporary astronomy are presented.  The first
comprises a major digital, blue-violet spectral classification program 
on Galactic O stars, that has already revealed three special categories 
or new members thereof: the Ofc class with C~III $\lambda$4650 emission
comparable to N~III $\lambda$4640, the Of?p class of magnetic oblique
rotators, and the ONn class of nitrogen-rich rapid rotators.  All of
these categories portend further understanding of massive stellar 
atmospheres and evolution.  The second example concerns the structure 
of massive young clusters and nebulae as a function of age on timescales 
of the order of or less than 10~Myr, which has provided a new insight 
into the Antennae major-merger starburst. 
\normalsize}
\end{minipage}

\section{The Morphological Process}

The morphological method in astronomy comprises the description of
unknown objects, in terms of certain well-defined criteria, differentially 
with respect to a reference frame of standard objects selected from the data 
themselves.  This approach will organize the empirical characteristics and 
often discover new ones, distinguish the normal majority from peculiar 
exceptions, and suggest or eliminate interpretational hypotheses.  An
essential element of the process is rigorous independence from external 
information, which ensures permanent validity of the description,
independently of errors in or revisions to the subsequent calibration and
interpretation.  Of course, morphology does not explain anything; rather,
it properly formulates the phenomena to be explained.

These concepts are displayed and interrelated graphically in the Walborn
Diagram (Figure~1) both in general astronomical terms and in those
specific to spectral classification and stellar astrophysics.  The key
point is the existence of an intermediate category, our Image of Nature,
between Nature itself and the ultimate objective of understanding Nature.
The role of morphology is to establish an Image that is sufficiently
accurate, complete, and unique to be usefully calibrated and interpreted.
Common errors are failure to recognize the existence or significance of
the intermediate category, confusion among the categories, and convolution 
of operations that should be sequential instead.  When the Image is
inadequate, no amount of physics will provide correct understanding.

\begin{figure}
\center
\includegraphics[width=\textwidth,viewport=0 270 590 570]{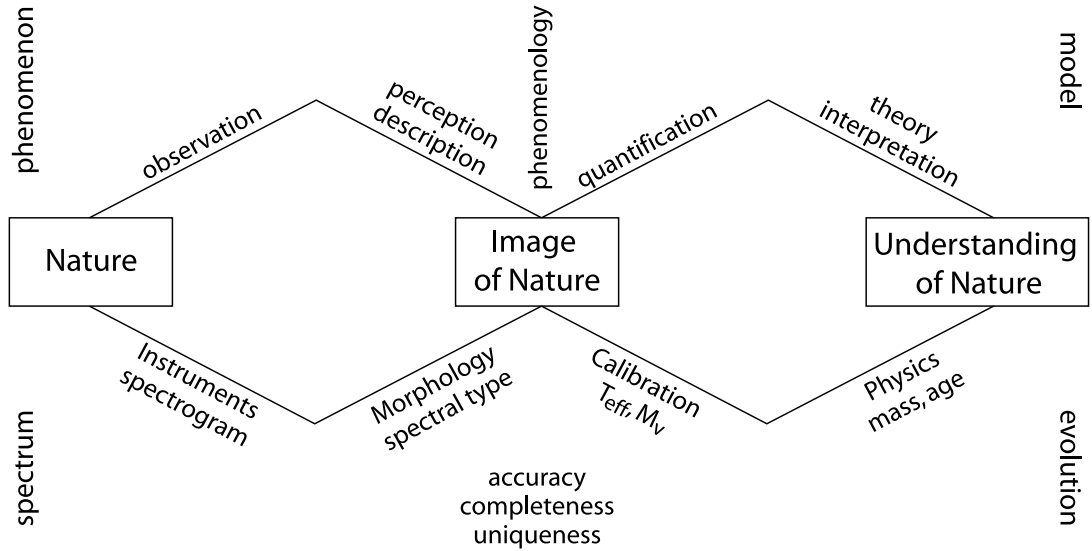}
\caption{\label{fig1} The Walborn Diagram, which displays the operations
linking Nature, our Image of Nature, and our understanding of Nature, in 
both general terms above and those specific to spectral classification or
stellar astrophysics below.}
\end{figure}

\section{OB Spectra}

Stellar spectral classification in the violet-blue-green optical wavelengths 
had its origins during the late 19th century, and it provided the
foundation of 20th century astrophysics.  Yet as illustrated in the 
following subsections, this technique continues to reveal new phenomena today.
The recent compendium Stellar Spectral Classification (Gray \& Corbally
2009) provides a definitive reference on the subject.  It provides a
valuable compilation of history, techniques, classification criteria,
and standard stars.  In particular, Chapter 3 on the OB stars presents
pertinent developments through the first decade of the 21st century,
including the remarkable correlations among the optical, ultraviolet,
infrared, and even X-ray spectra, which are leading current developments 
in many areas of astrophysics.

\subsection{GOSSS}

A major optical spectroscopic program led by J. Ma\'{\i}z Apell\'aniz at
the Instituto de Astrof\'{\i}sica de Andaluc\'{\i}a, the Galactic O-Star
Spectroscopic Survey (GOSSS), is acquiring moderate-resolution, high-S/N 
digital observations of all known Galactic O stars and many of type B 
to about 13th mag, about 2000 objects (Sota et al. 2011).  This unprecedented 
combination of quality, sample size, and homogeneity is improving the 
systematic and random accuracies of the classification, as shown in a new 
spectral atlas.  An expected corollary is the discovery of new objects and 
categories of special interest.  Three examples are described next.  

\subsection{Ofc Spectra}

Of spectra are defined by their selective emission lines of He~II
$\lambda$4686 and N~III $\lambda\lambda$4634-4640-4642; the gradual
development of these features is the basis of the O-type luminosity
classification.  A few otherwise normal Of supergiants with comparable
C~III $\lambda\lambda$4647-4650-4652 emission lines had been known, but
the GOSSS data show for the first time that this characteristic is
frequent near spectral type O5 at all luminosity classes and also increases 
gradually with luminosity (Walborn et al. 2010; Figure~2 here).  It
likely represents a new selective emission feature (Walborn 2001), i.e.
emission in particular transitions caused by anomalous level population
effects, which are very sensitive diagnostics of the atmospheric and wind
parameters (e.g., Corti, Walborn, \& Evans 2009).  Moreover, the C~III
emission appears to be present in some clusters and associations but not
others, possibly hinting at N/C abundance differences among them.  

\begin{figure}
\center
\includegraphics[width=0.5\textwidth,viewport=100 30 300 570]{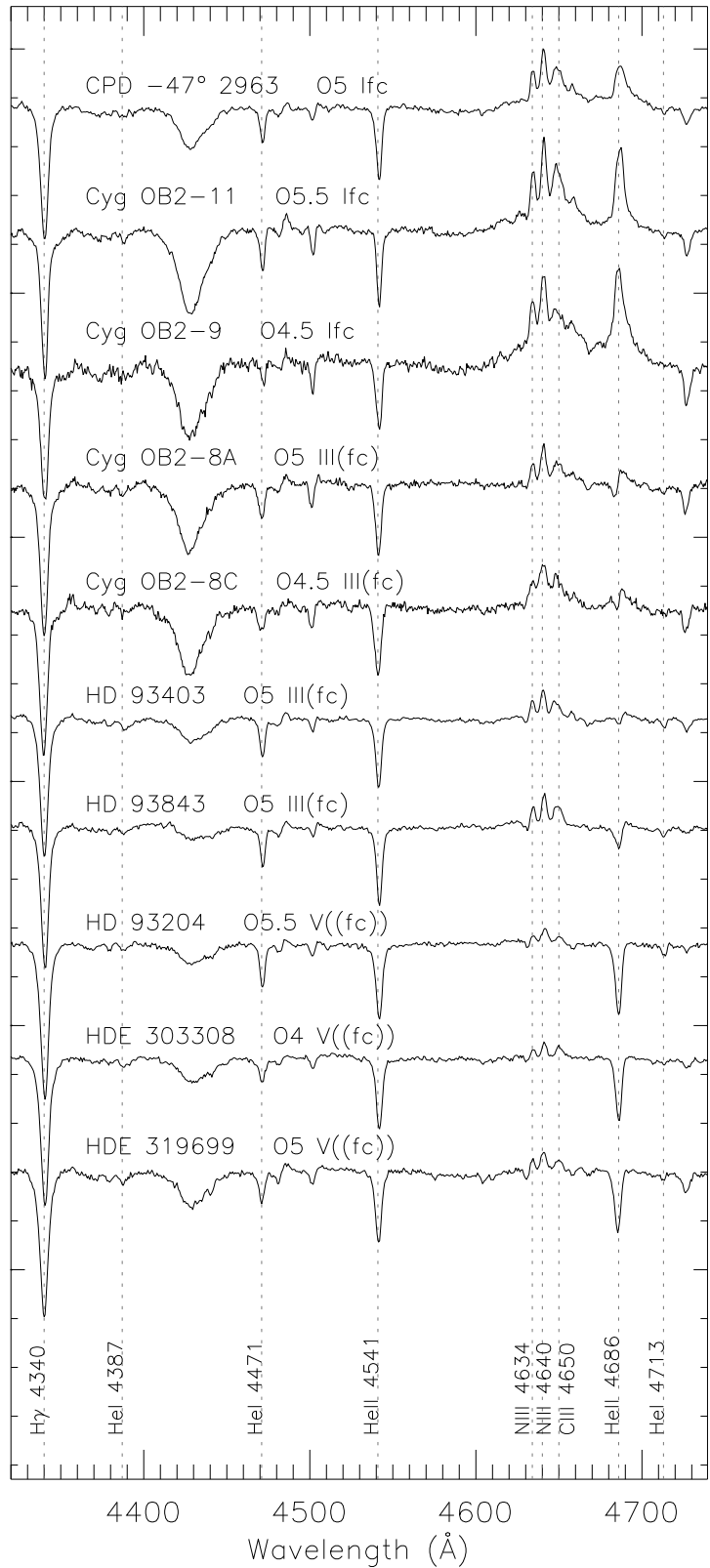}
\caption{\label{fig2} A luminosity sequence of Ofc blue-green spectra.
The longer ordinate tick marks are separated by 0.5 rectified continuum
intensity units.  From Walborn et al. (2010).}
\end{figure}

\subsection{Of?p Spectra}

The Of?p notation was introduced in the early 1970's to emphasize doubt
that three very peculiar and similar spectra corresponded to normal Of
supergiants.  In addition to C~III $\lambda\lambda$4647-4650-4652
emission lines comparable to N~III $\lambda\lambda$4634-4640-4642, they
showed evidence of circumstellar activity in terms of sharp absorption
features and P~Cygni profiles at H and He~I lines.  During the past
decade, all three have been shown to be extreme, periodic spectrum variables,
and soon thereafter their strong magnetic fields were detected (Martins et
al. 2010, Wade et al. 2011), revealing them as O-type magnetic oblique 
rotators.  GOSSS has already increased the membership of this rare category 
to five (Walborn et al. 2010; Figure~3 here); references to the prior 
developments can be found there.

\begin{figure}
\center
\includegraphics[width=0.6\textwidth,viewport=100 30 300 470]{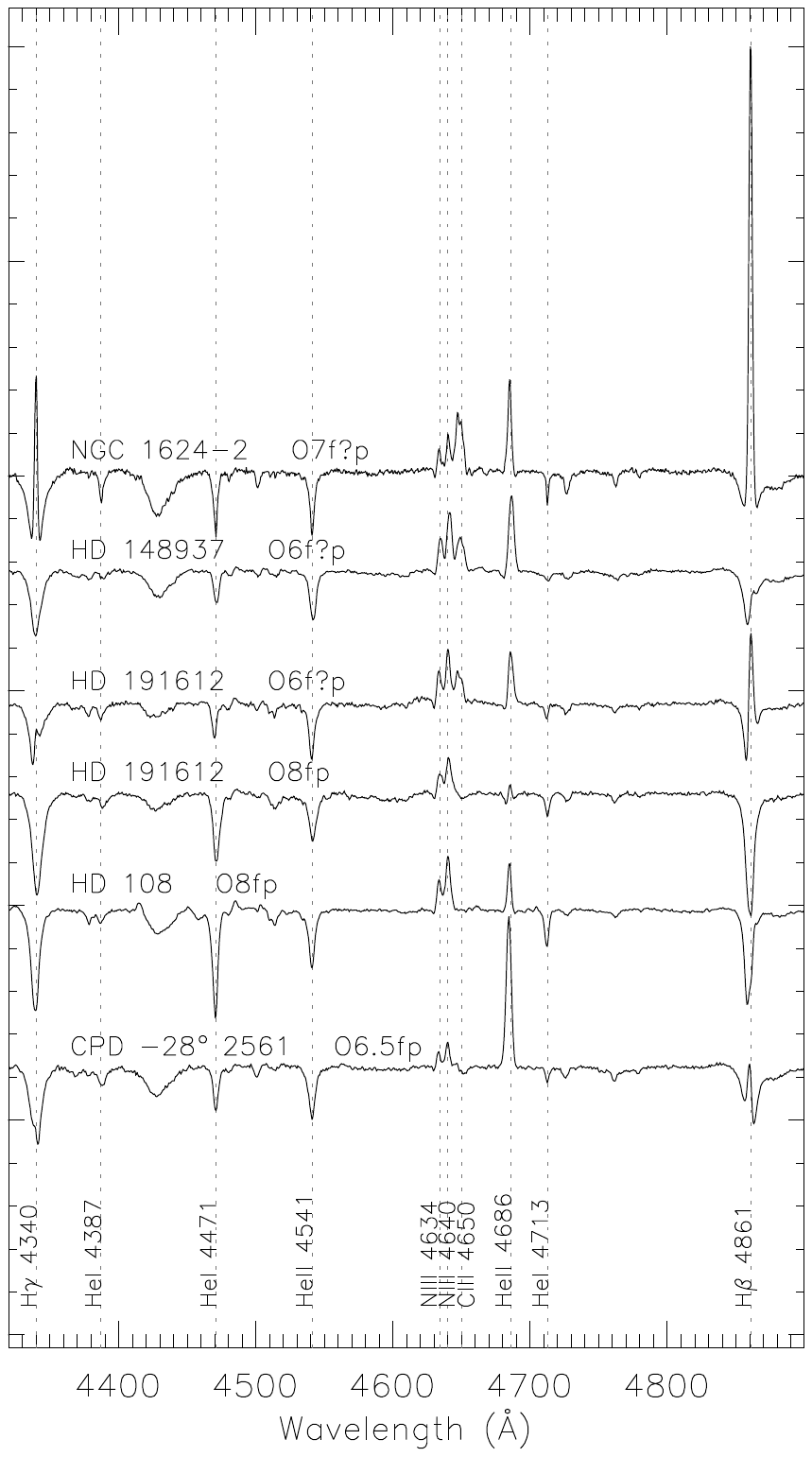}
\caption{\label{fig3} Of?p blue-green spectra.  HD~108, 148937, and 191612 
are the original members of the category; the last is shown in both
maximum and minimum states.  NGC~1624-2 was discovered to belong to this
class by GOSSS, while the membership of CPD~$-28^{\circ} 2561$ was first
established by the associated high-resolution southern survey led by
R.~Barb\'a and R.~Gamen (although a low-resolution observation is shown
here).  From Walborn et al. (2010).}
\end{figure}

\subsection{ONn Spectra}

The causes and effects of rotation in massive stellar evolution, involving
mixing or transfer of processed material to the surface and enhanced mass 
loss, are active areas of current research (Maeder \& Meynet 2000; Walborn 
2003; Langer et al. 2008).  In particular, enhanced N/C abundance ratios in 
the atmospheres and winds are a vital diagnostic of these effects.  The most 
rapidly rotating O star known, HD~191423, was shown to be nitrogen-enriched 
by Howarth \& Smith (2001).  GOSSS has already discovered two clones of this 
spectrum and confirmed several other less extreme but still rapidly rotating, 
late-O giants as members of this category (Walborn et al. 2011; Figure~3 here).
These spectra may well provide significant insights into the evolution of 
rotation and mixing between the main-sequence and supergiant stages, in the 
corresponding mass range. 

\begin{figure}
\center
\includegraphics[angle=90,width=1.1\textwidth,viewport=50 0 750 600]{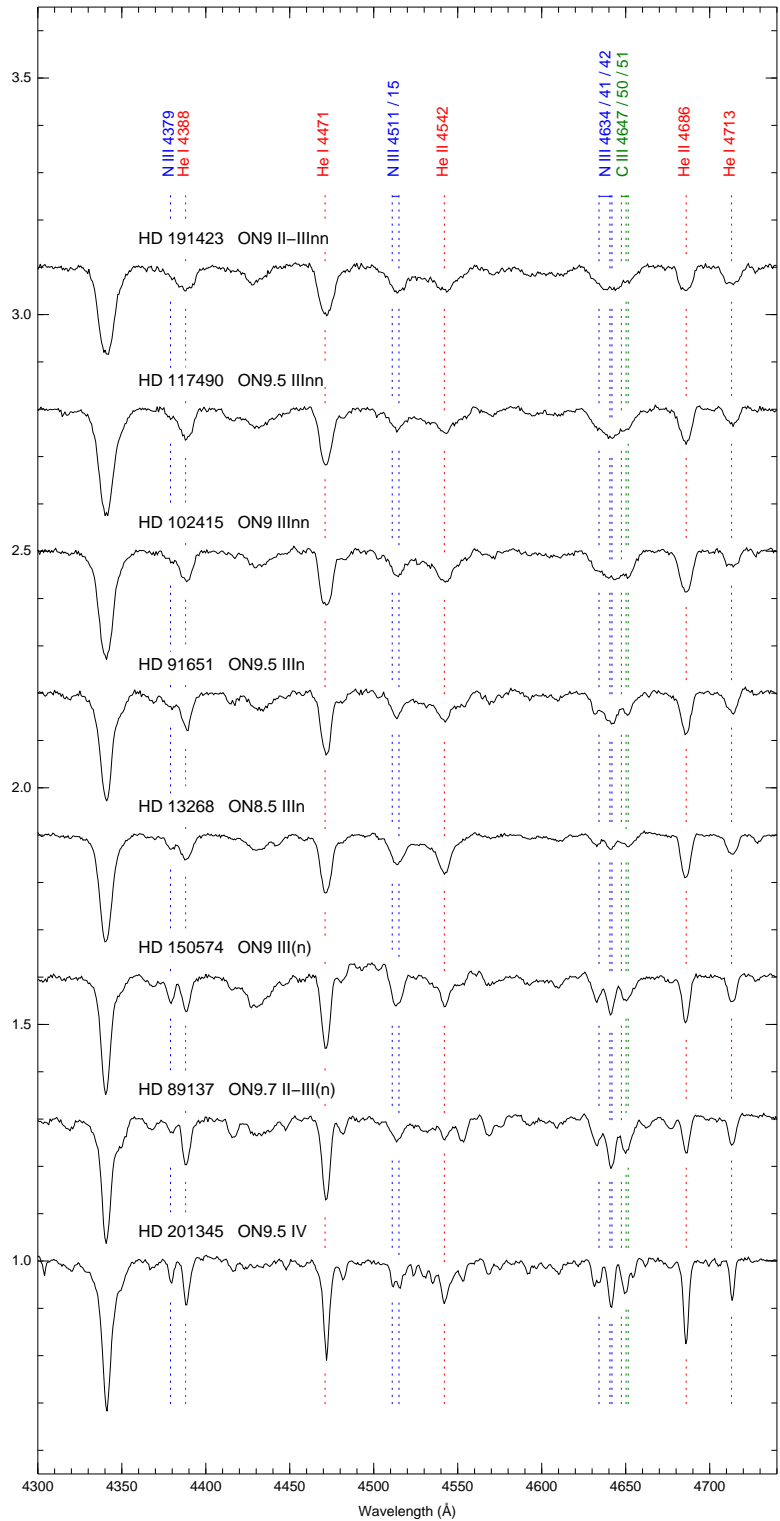} 
\caption{\label{fig4} A $v\sin i$ sequence of ONn blue-green spectra. 
All of these late-O giants (luminosity class III) and bright giants (II)
are strongly nitrogen-enhanced (``N'') and rapidly rotating (``n'').  The
approximate $v\sin i$'s corresponding to (n), n, and nn are 200, 300,
and 400 km~sec$^{-1}$, respectively.  HD~191423 is Howarth's Star, while
HD~102415 and 117490 are newly discovered as members of this category by
GOSSS.  The more slowly rotating subgiant (IV) HD~201345 is shown for
comparison.  To be further discussed by Walborn et al. (2011).}
\end{figure}

\section{OB Environments}

Because of the short timescale for massive stellar evolution, the
structure and content of the spectacular clusters and H~II regions in 
which most young massive stars are found also evolve qualitatively over the
first 10~Myr.  These evolving characteristics allow quite precise dating
of massive young clusters, to large distances in high-resolution images, 
from their morphology alone, as further discussed below. 

\subsection{Age Paradigms}

Table~1, from Walborn (2010), provides a standard sequence of young
clusters in terms of several well-defined parameters.  The latter are
derived from essentially complete (calibrated) spectral classification 
of the massive stellar content in these nearby Galactic objects.  
The degree to which all of these parameters are correlated in
coeval systems is remarkable.  As a result, objects that are too distant
for detailed study of their stellar content can be classified into these
evolutionary stages by their global morphologies alone.

\begin{table}[!ht]
\caption{Characteristics of massive young cluster age paradigms}
\smallskip
\begin{center}
{\small
\begin{tabular}{lllccll}
\hline\hline
\noalign{\smallskip}
\multicolumn{1}{c}{Object} &\multicolumn{1}{c}{Visually}
&\multicolumn{2}{c}{MS 
Turnoff} 
&Age &\multicolumn{1}{c}{H II,} &\multicolumn{1}{c}{Red}\\
\noalign{\smallskip}
\cline{3-4}
\noalign{\smallskip}
&\multicolumn{1}{c}{Brightest Stars} &\multicolumn{1}{c}{Spectrum}
&\multicolumn{1}{c}{Mass[$M_{\odot}$]} 
&[yrs] &\multicolumn{1}{c}{Dust} &\multicolumn{1}{c}{Sg}\\
\noalign{\smallskip}
\hline
\noalign{\smallskip}
Orion Nebula  &ZAMS O, (IR) &(PMS) & &$<10^6$           &Yes &No\\
Carina Nebula &O2, WNL    &O3 &\llap{1}00  &1--$2 \times 10^6$ &Yes &No\\
Scorpius OB1  &OB Sg       &O6 &50  &3--$4 \times 10^6$ &No  &No\\
Westerlund 1  &AF Sg        &O7--O8 &35   &4--$5 \times 10^6$ &No  &Yes\\
Perseus OB1   &AF Sg        &B0--B1 &20  &7--$9 \times 10^6$ &No  &Yes\\
\noalign{\smallskip}
\hline
\end{tabular}}
\end{center}
\end{table}
 
\subsection{Two-Stage Starbursts}

The next step in understanding massive young regions is the realization
that many or most of them consist of two phases from the preceding
sequence: an older first generation and a second, triggered generation at
the periphery, usually or always concentrated to one side (Walborn 2002).
(This latter characteristic possibly indicates that the initial collapse
is triggered externally near the surface, rather than globally toward the 
center of the precursor giant molecular cloud.)  
Thus, 30 Doradus in the Large Magellanic Cloud contains a Carina-phase
first generation and an Orion-phase second generation (Walborn et al.
1999a; Walborn, Ma\'{\i}z Apell\'aniz, \& Barb\'a 2002), while the 
second-ranked LMC H~II region Henize N11 is a giant shell with a
Scorpius~OB1-phase central association in a completely evacuated cavity and
Carina-phase nebulae at the periphery (Walborn \& Parker 1992; Walborn 
et al. 1999b).  In both cases the age {\it difference} between the two
generations is $\sim$2~Myr, but the absolute ages are greater in N11.
In the LMC objects, these conclusions are also based on complete spectral
classification of the massive stellar contents.  But then, the global
properties of similar regions in more distant galaxies can be used to
date them even though detailed spectral classification is not yet possible.  
E.g., NGC~604 in M33 is very similar to N11.  Even entire, recent 
star-formation subregions in some galaxies can be sequentially dated in this 
way, e.g. NGC~4214 (Mackenty et al. 2000), NGC~2363 (Drissen et al. 2000),
and NGC~6946 (Larsen et al. 2002).    

\subsection{The Antennae Revealed}

The Antennae, at a distance of about 20~Mpc, constitute the nearest major 
merger and are thus a key system for understanding this phenomenon, which
is ubiquitous at greater distances in the earlier Universe.  Accordingly,
it has been the subject of intensive investigation. With HST
imaging, Whitmore et al. (2010) have shown that it contains several 
hemispheric two-stage starbursts with global structures very similar to those 
of the nearby objects discussed above, despite their larger dimensions in the 
Antennae.  Three recent papers (Zhang et al. 2010, Karl et al. 2010, Teyssier 
et al. 2010) have established a timescale of order 10~Myr for the observed 
starburst, corresponding to the periastron passage in progress; Teyssier et al. 
even derive the essential result that the starburst is distributed rather
than nuclear.  The current interaction region of the two spiral disks has 
been identified with an extreme concentration of dust clouds and luminous 
infrared sources revealed by Spitzer (Wang et al. 
2004) and Herschel (Klaas et al. 2010).  
Nevertheless, the predominant structure of the global starburst in the
Antennae has not yet been clearly recognized.  It is readily discerned
by the above morphological procedures as a distorted Z-shaped age sequence 
(``The Mark of Zorro'') across the entire face of the system, avoiding 
the galactic nuclei, as shown in Figure~5.  The annotated ages are 
based on the structures (or absence) of the nebulae and clusters in 
the corresponding subregions, ranging from the pure dust/IR sources through, 
consecutively, centrally concentrated H~II regions, giant shells, and 
dissipating nebulosities, to a region of bright blue stars and possibly 
red supergiants without any nebulosity.  The intensity of the IR emission 
in the Spitzer images declines along the same sequence, with a single 
notable exception that may correspond to multiple generations at the site
of a giant shell (``S'' in the nomenclature of Whitmore et al.).  I propose 
that this sequence represents the track of the interaction region during the 
past 10~Myr or so.  This hypothesis will perhaps be verified by refined 
hydrodynamical modeling in the near future.

\begin{figure}
\center
\includegraphics[width=\textwidth]{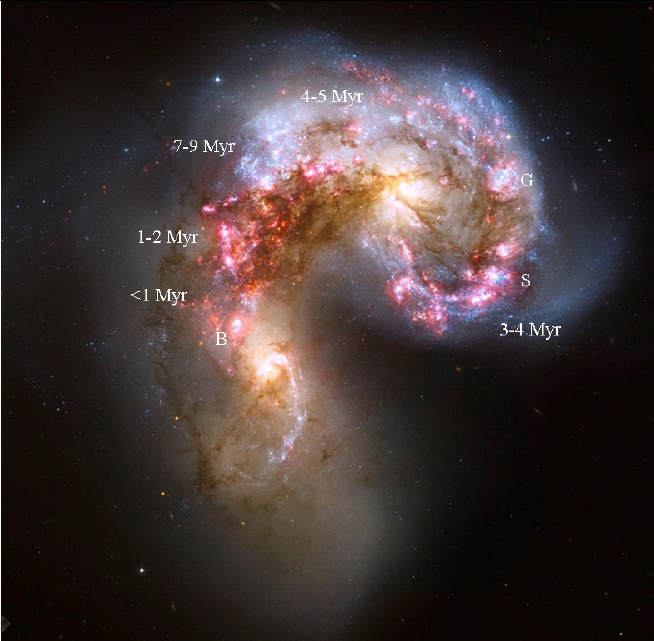}
\caption{\label{fig5} The Antennae major-merger starburst, as seen by the
HST Advanced Camera for Surveys.  The ages of various subregions
are marked based on the morphologies of the multiple nebulae and clusters
they contain; the most luminous Spitzer IR sources are located in the
youngest.  A monotonic, curving age sequence across the face of the system 
is revealed.  The letters are designations of particular complexes by
Whitmore et al. (2010).}
\end{figure}

\section{Concluding Remark}

It is evident from the results presented here that morphology remains a
vital discipline in astronomical research.

\small  
\section*{Acknowledgments}   

I thank the SEA~IX organizers for this kind invitation as well as generous 
travel and subsistence support.  Ancillary support was provided by NASA 
through grant GO-10898.01 from STScI, which is operated by AURA, Inc., 
under NASA contract NAS5-2655.

\end{document}